%% file: main.tex
\documentclass[sigconf]{acmart}
\input{datalabmacros} 
\input{individualmacros} %

\AtBeginDocument{%
  }

\setcopyright{acmlicensed}
\copyrightyear{2018}
\acmYear{2018}
\acmDOI{XXXXXXX.XXXXXXX}

\acmConference[Conference acronym 'XX]{Make sure to enter the correct
  conference title from your rights confirmation emai}{June 03--05,
  2018}{Woodstock, NY}
\acmISBN{978-1-4503-XXXX-X/18/06}

\begin{document}

\title{Fuzzy Integration of Data Lake Tables}

\author{Aamod Khatiwada}
\affiliation{%
  \institution{Northeastern University}
  \city{Boston}
  \country{USA}}
\email{khatiwada.a@northeastern.edu}

\author{Roee Shraga}
\affiliation{%
  \institution{Worchester Polytechnic Institute}
  \city{Worchester}
  \country{USA}}
\email{rshraga@wpi.edu}

\author{Ren\'ee J. Miller}
\affiliation{%
  \institution{Northeastern U. \& U. of Waterloo}
  \city{Boston}
  \country{USA}}
\email{miller@northeastern.edu}

\renewcommand{\shortauthors}{Khatiwada et al.}

\include{sections/abstract}


\maketitle
\input{sections/introduction}
\input{sections/related_work}

\input{sections/system_description}
\input{sections/experiments}
\input{sections/conclusion}
\bibliographystyle{ACM-Reference-Format}
\bibliography{main}

\end{document}

%% file: datalabmacros.tex
\usepackage{amsmath}
	\usepackage{cancel}
\usepackage{mathtools}
\usepackage{graphicx} 
\usepackage{url}
\usepackage{enumitem} 

\usepackage{stmaryrd} 

\usepackage{upgreek} 
\usepackage{bbding} 

\usepackage{listings} 
\usepackage{tabularx} 
\usepackage{booktabs} 
\usepackage{multirow}
\usepackage{hhline} 
\usepackage{diagbox}
\usepackage{pgfplotstable} 

\usepackage{xcolor,colortbl}    

\usepackage{tikz} 
\usetikzlibrary{matrix}
\usetikzlibrary{calc}
\usetikzlibrary{math}
\usetikzlibrary{arrows.meta,positioning}
\usetikzlibrary{intersections, pgfplots.fillbetween}

\usepackage{easybmat} 

\usepackage{adjustbox}
\def\eox{\unskip\kern 10pt{\unitlength1pt\linethickness{.4pt}$\diamondsuit${}}} 

\usepackage{rotating}		

\usepackage{xspace} 

\usepackage{subcaption} 
\newcommand{\hide}[1]{}

\usepackage[ruled,noend,linesnumbered]{algorithm2e} 

\usepackage{setspace} 

\DontPrintSemicolon     
\SetNlSty{}{}{}                
\SetAlgoInsideSkip{smallskip}   

\SetAlFnt{\small}			
\SetAlCapFnt{\small}		
\SetAlCapNameFnt{\small}

\usepackage[capitalise,nameinlink]{cleveref} 

\crefname{section}{Sec.}{Secs.}
\crefname{example}{Ex.}{Exes.}

\hypersetup{                                                            
	pdfpagemode=UseNone,               
    pdfstartview=Fit,
    pdfpagelayout=SinglePage,       
    bookmarks=false,
    bookmarksnumbered=true,
    bookmarksopen=false,             
    pdftex=true,
    unicode,
    colorlinks=true,       
    linkcolor=magenta,          
    citecolor=cyan,  
    filecolor=magenta,      
    urlcolor=blue           
}

\usepackage{aliascnt}  	

\newaliascnt{corollary}{theorem}

\aliascntresetthe{corollary}

\newaliascnt{example}{theorem}
\newtheorem{example}[example]{Example}
\aliascntresetthe{example}

\newaliascnt{definition}{theorem}
\newtheorem{definition}[definition]{Definition}
\aliascntresetthe{definition}

\newaliascnt{proposition}{theorem}

\aliascntresetthe{proposition}

\newaliascnt{lemma}{theorem}

\aliascntresetthe{lemma}

\newaliascnt{conjecture}{theorem}

\aliascntresetthe{conjecture}

\newtheorem{questionW}{Question}
\newtheorem{resultW}{Result}

{\end{itshape}
}
\AtEndPreamble{%
    \hypersetup{colorlinks,
      linkcolor=purple,
      citecolor=blue,
      urlcolor=ACMDarkBlue,
      filecolor=ACMDarkBlue}}

\usepackage{tcolorbox}		
\tcbuselibrary{breakable,skins}		

\tcbset{examplestyle/.style={
		enhanced jigsaw,	
		colback=blue!08,	
		colframe=blue!08,	
		arc=2mm,
		boxrule=0pt,		
		left=1mm,
		right=1mm,
		left skip= 0mm,  
		right skip= 0mm, 
		top=1mm,		
		bottom=1mm,		
		breakable,		
		parbox = false,		
		before={\par\pagebreak[0]\vspace{1mm}\parindent=0pt},		
		after={\par\pagebreak[0]\vspace{1mm}\parindent=0pt},				
		bottomrule = 0mm,
		boxsep = 0mm,					
		topsep at break=0pt,			
		bottomsep at break=0pt,			
		pad at break=0mm,
		pad before break=1mm,		
		pad after break=1mm,		
		bottomrule at break=0mm,
		toprule at break=0mm,		
		}}

\usepackage{footnote}

\tcolorboxenvironment{example}{examplestyle}

\makeatletter
\DeclareRobustCommand*\uell{\mathpalette\@uell\relax}
\newcommand*\@uell[2]{
  \setbox0=\hbox{$#1\ell$}
  \setbox1=\hbox{\rotatebox{10}{$#1\ell$}}
  \dimen0=\wd0 \advance\dimen0 by -\wd1 \divide\dimen0 by 2
  \mathord{\lower 0.1ex \hbox{\kern\dimen0\unhbox1\kern\dimen0}}
}
\makeatother

\usepackage{marginnote}

\newcommand{\introparagraph}[1]{\textbf{#1.}} 

\renewcommand{\epsilon}{\varepsilon} 

\usepackage{scalerel}
\usepackage{tikz}
\usetikzlibrary{svg.path}

\definecolor{orcidlogocol}{HTML}{A6CE39}
\tikzset{
  orcidlogo/.pic={
    \fill[orcidlogocol] svg{M256,128c0,70.7-57.3,128-128,128C57.3,256,0,198.7,0,128C0,57.3,57.3,0,128,0C198.7,0,256,57.3,256,128z};
    \fill[white] svg{M86.3,186.2H70.9V79.1h15.4v48.4V186.2z}
                 svg{M108.9,79.1h41.6c39.6,0,57,28.3,57,53.6c0,27.5-21.5,53.6-56.8,53.6h-41.8V79.1z M124.3,172.4h24.5c34.9,0,42.9-26.5,42.9-39.7c0-21.5-13.7-39.7-43.7-39.7h-23.7V172.4z}
                 svg{M88.7,56.8c0,5.5-4.5,10.1-10.1,10.1c-5.6,0-10.1-4.6-10.1-10.1c0-5.6,4.5-10.1,10.1-10.1C84.2,46.7,88.7,51.3,88.7,56.8z};
  }
}

\RequirePackage{etoolbox}
\DeclareRobustCommand\orcidicon[1]{\href{https://orcid.org/#1}{\mbox{\scalerel*{
\begin{tikzpicture}[yscale=-1,transform shape]
\pic{orcidlogo};
\end{tikzpicture}
}{|}}}}

%% file: individualmacros.tex
\usepackage{array}
\usepackage{makecell}



%% file: sections/abstract.tex
\begin{abstract}
Data integration is an important step in any data science pipeline where the objective is to unify the information available in different datasets for comprehensive analysis. Full Disjunction, which is an associative extension of the outer join operator, has been shown to be an effective operator for integrating datasets. It fully preserves and combines the available information. Existing Full Disjunction algorithms only consider the equi-join scenario where only tuples having the same value on joining columns are integrated. This, however, does not realistically represent an open data scenario, where datasets come from diverse sources with inconsistent values (e.g., synonyms, abbreviations, etc.) and with limited metadata. So, joining just on equal values severely limits the ability of  Full Disjunction to fully combine datasets. Thus, in this work, we propose an extension of Full Disjunction to also account for ``fuzzy'' matches among tuples. We present a novel data-driven approach to enable the joining of approximate or fuzzy matches within Full Disjunction. Experimentally, we show that fuzzy Full Disjunction does not add significant time overhead over a state-of-the-art Full Disjunction implementation and also that it enhances the integration effectiveness.
\end{abstract}

%% file: sections/introduction.tex
\section{Introduction}
\begin{figure}
  \includegraphics[scale = 0.48]{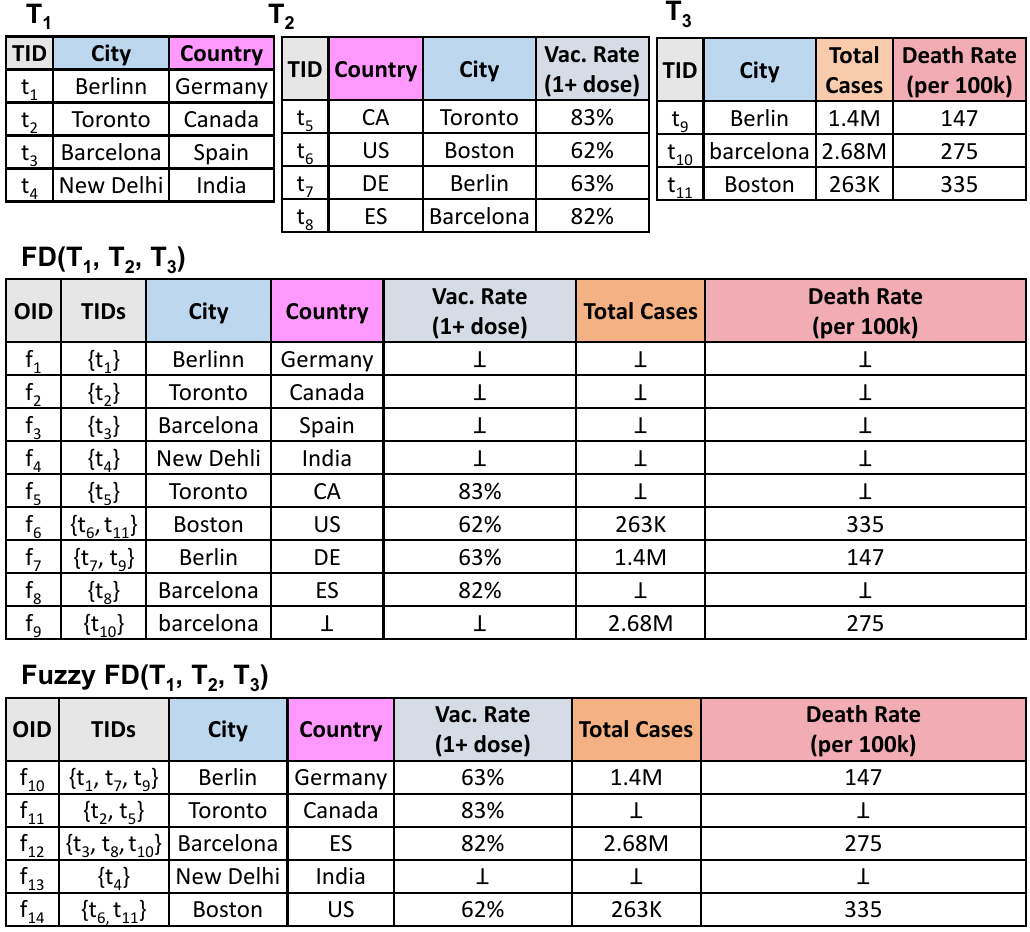}
  \caption{Tables about COVID-19 cases in different cities. The column headers are given only for easy reference. However, they may not be available in practice. 
  }
\label{fig:main_example}
\end{figure}
Data lakes store an enormous amount of heterogeneous data. Within data lakes, tables (e.g., CSV files) are one of the most prevalent data formats~\cite{2019_ravat_data_lakes_trends, 2023_hai_datalake_survey}, and tables are useful for data scientists to run various analyses and make decisions~\cite{2019_nargesian_data_lake_management, 2020_galhotra_s3d}. 
However, diverse tables covering information from different topics may use inconsistent values (e.g., synonyms or abbreviations) and may have unreliable metadata (e.g., table names and column headers) making it difficult for data scientists to find  data lake tables that are relevant for their analysis. 
Consequently, different semantic table search techniques have been proposed~\cite{2018_fernandez_aurum, 2018_nargesian_tus, khatiwada2024tabsketchfm}. 
Such techniques generally input keywords~\cite{2021_ouellette_ronin, 2019_Brickley_google_dataset_search,2018_fernandez_aurum} or an existing table as query~\cite{2023_fan_starmie,2018_nargesian_tus, 2019_zhu_josie} and search for  data lake tables that are most relevant (e.g., unionable~\cite{2023_khatiwada_santos, 2018_nargesian_tus, 2023_hu_autotus} or joinable~\cite{2021_dong_efficient_joinable_table_discovery, DBLP:journals/pvldb/ZhuNPM16}) to the query.

After discovery, the required information for analysis could be scattered among query and searched tables. So, the next natural step 
is integration
and the generation of a unified view of relevant data~\cite{2022_khatiwada_alite, 2009_bleiholder_data_fusion}. 
Two major challenges have been considered for integrating a set of discovered tables.
The first is to determine which columns should be aligned together in the integrated table. 
A possible solution could be to align the columns having the same column headers. However, 
since data lake tables may have missing, inconsistent, and unreliable column headers, this becomes challenging.
Hence, we cannot rely on them for comprehensive integration and to make them consistent, techniques such as schema matching are applied~\cite{koutras2021valentine, 2009_bleiholder_data_fusion}. 

After determining the aligning columns, the second challenge is to find an integration operator (or query) to merge the tuples and generate an integrated table.
Basic integration operators such as inner join, union, outer join, and so on, may not be effective as they may not retain all the information during integration~\cite{2003_kanza_computing_FD, 1994_legaria_outerjoin_as_disjunctions, 2019_paganelli_parallelizing_fd}. For instance, the inner join operator, when integrating a set of tables (an \emph{integration set}), does not retain a tuple if it has no joining partner even in a single table in a large integration set. Outer join solves the inner join problem by retaining  tuples without join partners. However, the outer join is not an associative operator and different orders of applying outer join over a set of tables generate different sets of partially integrated tuples~\cite{1996_ullman_outer_join_gamma_cycles,2006_cohen_poly_delay_fD}.
Consequently, ~\citet{1994_legaria_outerjoin_as_disjunctions} introduced the Full Disjunction (FD) operator, which is an associative version of the outer join operator. FD joins each tuple in the tables to be integrated in a maximal way such that each tuple is represented and no tuples remain incomplete in the integrated table~\cite{1996_ullman_outer_join_gamma_cycles}.
Hence, FD has been considered an optimal way of integrating information present in different tables~\cite{1996_ullman_outer_join_gamma_cycles}. We refer to the literature for further details on Full Disjunction~\cite{1996_ullman_outer_join_gamma_cycles, 2006_cohen_poly_delay_fD}, including its scalable~\cite{ 2022_khatiwada_alite} and parallelized implementations ~\cite{2019_paganelli_parallelizing_fd}.

Notice however that the existing Full Disjunction algorithms only consider joining tuples on equal values~\cite{1994_legaria_outerjoin_as_disjunctions, 2006_cohen_poly_delay_fD, 2022_khatiwada_alite}.
In reality, the data lake tables come with inconsistencies such as abbreviations, synonyms, and more. So, relying on equal value joins impacts Full Disjunction's ability to  integrate the tables well and also impacts the usage of integrated tables for 
 downstream tasks. Instead we need "fuzzy" matching between the values.
Therefore, we propose an extension of the Full Disjunction (FD) operator that also accounts for fuzzy matches between the values. Specifically, 
our solution first resolves the inconsistency between the column cells representing the same values. After making the values consistent, it applies the FD operator to integrate the tuples.

\begin{example}
    Consider Tables $T_1$, $T_2$ and $T_3$ about COVID-19 cases in~\cref{fig:main_example}.
    The columns $TID$ and $OID$ are used for illustration to clearly indicate which tuples (in $TID$ column) were integrated to produced this new tuple ($OID$ column).
    For simplicity, columns that align are given the same name in the three tables
    and are highlighted in the same colors. 
    Table $FD(T_1, T_2, T_3)$ shows the Full Disjunction result using an equi-join. 
    Since $T_1$ has a typo in Tuple $t_1$ (Berlinn), Full Disjunction does not integrate it with other tuples about Berlin ($t_7$ and $t_9$) forming separate tuples $f_1$ and $f_7$. 
    Furthermore, as two aligning \texttt{Country} Columns in Tables $T_1$ and $T_2$ contain the full names and codes of countries respectively, FD does not integrate Tuples $t_2$ and $t_5$ and Tuples $t_3$ and $t_8$. Moreover, tuples $t_3$ and $t_{10}$ are both about Barcelona but they are not integrated by FD as they are represented in different cases (Barcelona in $t_3$ and barcelona in $t_{10}$). 
    On the other hand, Fuzzy FD($T_1$, $T_2$, $T_3$) shows tuples integrated using our proposed algorithm where the tuples are integrated maximally without redundancy. 
\end{example}
Next, we summarize our contributions.
\begin{itemize}
    \item \textbf{Fuzzy Full Disjunction:} To the best of our knowledge, we are the first to propose fuzzy integration of the tuples using the Full Disjunction Operator. Specifically, our method first determines the fuzzy matches, makes them consistent, and then applies Full Disjunction.
    \item \textbf{Empirical Evaluation:} We show experimentally that our novel Fuzzy Full Disjunction method, without significantly increasing runtime, enhances integration effectiveness.

\end{itemize}

%% file: sections/related_work.tex
\introparagraph{Related Work}
\citet{1994_legaria_outerjoin_as_disjunctions} introduced FD as an associative alternative to the outer join operator which is computed by applying outer join in all possible orders of the input tables. Next, the results are outer unioned to generate all possible FD tuples. A subsumption operator is then applied to eliminate a tuple that is contained in another tuple (i.e., tuples containing partial information). \citet{1996_ullman_outer_join_gamma_cycles} state that FD is the right semantics for data integration and several others~\cite{2003_kanza_computing_FD, 2006_cohen_poly_delay_fD, 2019_paganelli_parallelizing_fd} propose algorithms to compute FD faster in practice.
Recently, \citet{2022_khatiwada_alite} proposed ALITE, which uses the Full Disjunction operator to integrate  data lake tables discovered using different table search techniques. Since data lake tables have inconsistent and unreliable column headers~\cite{2018_miller_open_data_integration, 2023_hai_datalake_survey}, ALITE first
determines the matching columns in the tables to be integrated by applying holistic schema matching~\cite{su2006holistic} over  column-based pre-trained embeddings. After that, ALITE uses the (Natural) Full Disjunction~\cite{1994_legaria_outerjoin_as_disjunctions}, 
over the matched columns to produce an integrated table.
However, unlike this work, all the prior work considers equi-join integration of tuples, i.e., the values of the tuples to be integrated are consistent and can be matched using the equality operator. We relax this assumption and propose a novel way of applying Full Disjunction over  data lake tables using approximate or fuzzy joins.

Previous work has addressed the challenge of identifying fuzzy inner joins.
~\citet{2017_zhu_autojoin} determined fuzzy matches by employing string transformations such as matching n-grams of cell values and concatenating cell values. ~\citet{2021_li_auto_fuzzy_join} determined fuzzy matches between the values within a column pair by selecting suitable parameters for the given input table. They framed fuzzy matching as an optimization task with an objective of maximizing matching recall under a precision constraint. Different from our work, they focus on  fuzzy matching between a pair of columns, whereas our task requires matching values across a set of columns.

%% file: sections/system_description.tex
\section{System Description}
\label{section:system_description}

\begin{figure}
  \includegraphics[scale = 0.33]{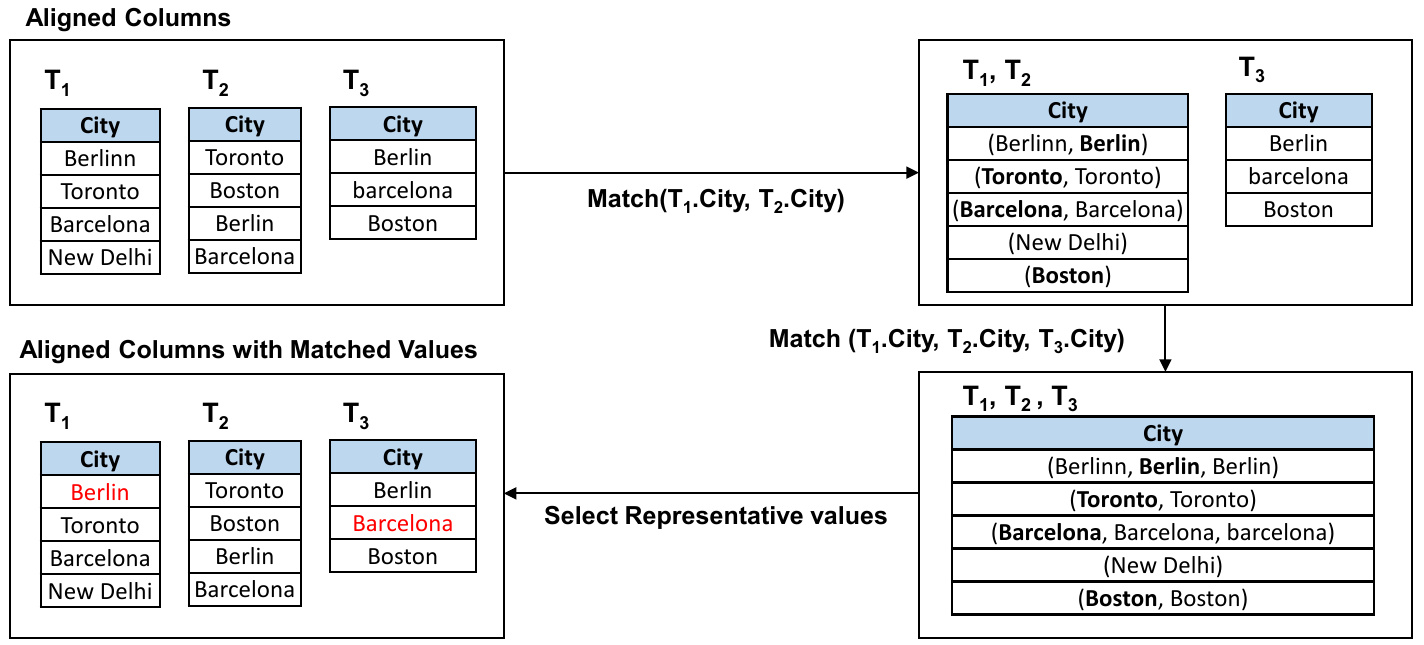}
  \caption{Applying Match Values component over the aligning columns for Fuzzy Integration of Data Lake Tables.}
\label{fig:matching_example}
\end{figure}

Our fuzzy Full Disjunction
processes a set of input tables with aligned columns and outputs an integrated table. Over the aligned columns, we apply value matching
which addresses inconsistencies among join values.  
Once 
value matches are identified, the matching values are replaced with a single consistent value before applying the FD operator. 
Here we discuss the implementation of the value matching.
For details on Aligning columns and Full Disjunction operator, we refer to the prior work on table integration~\cite{2022_khatiwada_alite}.
\subsection{Value Matching}
Let $T_l$ be a table in the integration set.
We denote a set of aligning columns using $C$ and $i_{th}$ column in the set using $c_i$.  Similarly, we represent $j_{th}$ value in a list of values $V$ using $v_j$.
Furthermore, $c_i.v_j$ represents the $j_{th}$ value of Column $c_i$. 
As the columns from the same table do not align, there can be at most one column from a table in a set of aligning columns.
Following the literature on the problem of entity matching (EM)~\cite{2020_li_ditto}, we consider a clean-clean value matching scenario~\cite{2021_christophides_er_survery}. i.e., the inconsistency between the values within a column is resolved, and we want to match them across the columns. 
Note that, different from the EM problem, here we consider single values for matching rather than full rows (tuples). Nevertheless, we will experiment with EM as a post-integration task in~\cref{section:experiments}.
Next, we formally define the Fuzzy Value Match problem.


\begin{definition}[Fuzzy Value Match Problem]
    Given a set of aligning columns $c_1$, $c_2$ $\dots$ $c_n$, a list of their values $[c_i.v_j | i \in 1 \dots n; j \in 1, 2, \dots]$, 
    a matching threshold $\theta$,
    and a distance function $dist(.)$,
    The Fuzzy Value Match Problem is finding a disjoint set of values $V_1$, $V_2$, $\dots$ $V_k$ such that 
    $\forall u, v \in V_i$ ($1\leq i \leq k$) input,
$dist(u,v) < \theta$.
\end{definition}



\subsection{Match Values Implementation}
Now, we explain how we implement the Match Values component that determines the fuzzy matches between the values.

\noindent\textbf{Embed Column Values: }
We first represent each 
column value in a fixed-dimension embedding space to ensure that  matched values are close to each other in this embedding space. For instance, cells referring to Canada with values "Canada" and "CA" in $T_1$ and $T_2$~(\cref{fig:main_example}), respectively, are embedded near each other, while "Germany" and "CA" are embedded farther apart. Similarly, we aim to embed "Berlinn" and "Berlin" close to each other as they both represent Berlin.
In our system, we embed each cell using Mistral-7B-Instruct model,
\footnote{\url{https://huggingface.co/docs/transformers/main/en/model_doc/mistral}}
a recent large language model that we used in our experimental analysis (detailed in~\cref{section:experiments}).


\noindent\textbf{Determine Fuzzy Matches: }
In the context of the clean-clean scenario~\cite{2021_papadakis_er_tutorial, 2021_christophides_er_survery}, 
the values within each column are consistent (that is, two values have the same meaning iff they are identical). In what follows, our approach identifies fuzzy matches between values across aligning columns. We initiate this process by selecting a pair of aligning columns and determining the fuzzy matches between their respective sets of values.
To determine these matches, we compute the cosine distances between the embeddings of cell values from the first column and those from the second column. 
Based on these distances, we perform bipartite matching between the values of the column pairs. 
Specifically, we apply  linear sum assignment algorithm~\cite{2016_crouse_assignment_algorithm} that identifies an optimal bipartite match between the values, minimizing the total distance between the matched values.  
Note that we do not allow matches whose distance is higher than the threshold $\theta$.


\begin{example}
    Consider the \texttt{Country} columns of Tables $T_1$ and $T_2$ in \cref{fig:main_example} that are aligned. We apply bipartite matching between their sets of values. Based on the embeddings, Germany is matched with DE, Canada is matched with CA, and Spain is matched with ES. Bipartite matching matches India in $T_1$ with US in $T_2$ but their match score is above the threshold. So, this match is discarded and these values are placed in separate value sets.
\end{example}

Once we determine a match between the values of two columns, we outer join the columns to generate a combined column.
If a value in one column is not matched with a value in another column, it is left in a singleton set represented by its embedding.  
If two values match, we select the most representative value embedding, i.e., the one that appears most frequently in the list of all values from the aligning columns. In the case of a tie, we select a value from the first table among the two matching tables each time, to keep the assignment consistent.
This process produces a combined column, which we then use for bipartite matching with another aligning column. We continue producing the combined column and matching it with another aligning column until all fuzzy matches in the set of aligning columns are determined.

\begin{example}
\label{example:matching_example}
    \Cref{fig:matching_example} illustrates three aligning \texttt{City} columns from $T_1$, $T_2$, and $T_3$ in \cref{fig:main_example}. In the first step, we match the \texttt{City} columns from $T_1$ and $T_2$. This results in Berlinn, Toronto, and Barcelona from $T_1$ being matched with Berlin, Toronto, and Barcelona from $T_2$, respectively. New Delhi remains unmatched.
Since Berlin appears twice and Berlinn appears once across all three columns, we select Berlin for the combined column (bold in Table $T_1, T_2$ on the top-right corner). For the other two values (Toronto and Barcelona), as they are identical, we simply select one for the combined column. 
New Delhi with no other matches is also added to the combined column.
Next, we match the combined column from $T_1$ and $T_2$ with the \texttt{City} column from $T_3$. This process results in the final combined column containing the values Berlin, Toronto, Barcelona, New Delhi, and Boston (bottom-right corner).
\end{example}

The column values in the final combined column are selected as the representative values for each set of matched values. Then, we replace all of the values across the aligning columns with their respective representative value. For instance, continuing~\cref{example:matching_example}, as shown in the bottom-left corner of~\cref{fig:matching_example}, we replace Berlinn and barcelona with their representative values Berlin and Barcelona respectively.

After matching the values in each set of aligning columns, all value level inconsistency gets resolved. Consequently, we apply the equi-join Full Disjunction operator and it integrates the tuples without missing the integration on fuzzy values. In our implementation, we use the state-of-the-art Full Disjunction algorithm~\cite{2022_khatiwada_alite}.

%% file: sections/experiments.tex
\section{Experiments}
\label{section:experiments}
\begin{table}
\setlength{\tabcolsep}{3pt}
\centering
\caption{Value Matching effectiveness of different models in Auto-Join Benchmark. The best score along each column is in bold; the second best score is \underline{underlined}.}
  \label{tab:fuzzy_matching_effectiveness}
  \begin{tabular}{l|rrr|}
    \toprule
    \textbf{Model}&{\textbf{Precision}}&{\textbf{Recall}}&{$\mathbf{F_1}$\textbf{-Score}}\\
    \hline
    FastText&0.70&0.67&0.66\\
    BERT&0.72&0.76&0.73\\
    RoBERTa&0.73&0.77&0.74\\
    Llama3&\underline{0.81}&\underline{0.85}&\underline{0.81}\\
    Mistral&\textbf{0.81}&\textbf{0.86}&\textbf{0.82}\\
    
  \bottomrule
\end{tabular}
\end{table}

    

Now we empirically evaluate our Fuzzy Full Disjunction method with an equi-join-based FD algorithm.

\subsection{Experimental Setup}
We run all our experiments using Python 3.10 on a server having Intel(R) Xeon(R) Gold 6346 CPU @ 3.10GHz and NVIDIA A40 GPU.
Through experiments, we intend to answer the following:
(i) How effective is our method in determining the fuzzy matches between the values?
(ii) How efficient is our proposed method in integrating the tables considering fuzzy matches?

We implement Full Disjunction operator~\cite{2022_khatiwada_alite} using the publicly available code.\footnote{\label{footnote:alite_github}\url{https://github.com/northeastern-datalab/alite}}
Furthermore, we use scipy's implementation of the linear sum assignment algorithm to perform bipartite matching.\footnote{\url{https://docs.scipy.org/doc/scipy/reference/generated/scipy.optimize.linear_sum_assignment.html}}
Following the literature~\cite{2018_nargesian_tus, 2021_dong_efficient_joinable_table_discovery}, we report the results with the matching threshold $(\theta)$ of 0.7, which gives the best results. 

\introparagraph{Benchmarks}
We run our experiments over publicly available join benchmarks from the literature.

\textit{(i) Auto-Join Benchmark.}
\textbf{Auto-Join} is a publicly available\footnote{\url{https://github.com/Yeye-He/Auto-Join}}
fuzzy tuple matching Benchmark that comes with 31 integration sets of tables covering 17 topics such as songs, government official details, and so on~\cite{2017_zhu_autojoin}. Each integration set contains sets of aligning columns (having around 150 values per column on average) that can be joined in a fuzzy manner considering clean-clean scenario~\cite{2021_christophides_er_survery}. We run our experiments over such joining columns.  


\textit{(ii) ALITE Benchmark.} 
~\citet{2022_khatiwada_alite} created a set of datasets using open data tables to evaluate the effectiveness and efficiency of table integration methods. We use their Entity Matching Dataset to study effectiveness. Furthermore, they also created a benchmark based on \textbf{IMDB} movie dataset containing about 106M tuples distributed in 6 tables, to study the efficiency of regular Full Disjunction operator.\footnote{\url{https://datasets.imdbws.com/}}
Specifically, they sampled the rows from the IMDB tables to create integration sets containing 5K to 30K input tuples and study FD's runtime over them. Although this is an equi-join benchmark, as the Match Values component still needs the same time to check for the fuzzy matches even if they do not exist, we use this benchmark to study the efficiency of our method against the baselines.

\introparagraph{Baselines}
To embed the column values, we evaluate different embedding baselines. Specifically, we implement a publicly available word embedding model (\textbf{FastText}~\cite{2016_joulin_fasttext}).\footnote{\url{https://github.com/facebookresearch/fastText}} We also implement two pre-trained language models (\textbf{BERT}~\cite{2019_devlin_bert} 
and \textbf{RoBERTa}~\cite{2019_liu_roberta}) and two recent large language models: \textbf{Mistral} (Mistral-7B-Instruct-v0.3) and \textbf{Llama3} (Meta-Llama-3-8B-Instruct) available in Hugging Face library.\footnote{\url{https://huggingface.co/}} For each language model, we pass the cell values through its layers and extract the embeddings of the last hidden layer. 
Moreover, we use \textbf{ALITE}, the state-of-the-art table integration system based on full disjunction, as an integration baseline.$^{\ref{footnote:alite_github}}$

\introparagraph{Evaluation Metrics}
To report the value match effectiveness, we use the standard Precision ($P$), Recall ($R$), and $F_1$-Score ($F_1$). 
We also report runtime to compare efficiency. 


\subsection{Results}
\label{section:results}
\begin{figure}
  \includegraphics[scale = 0.33]{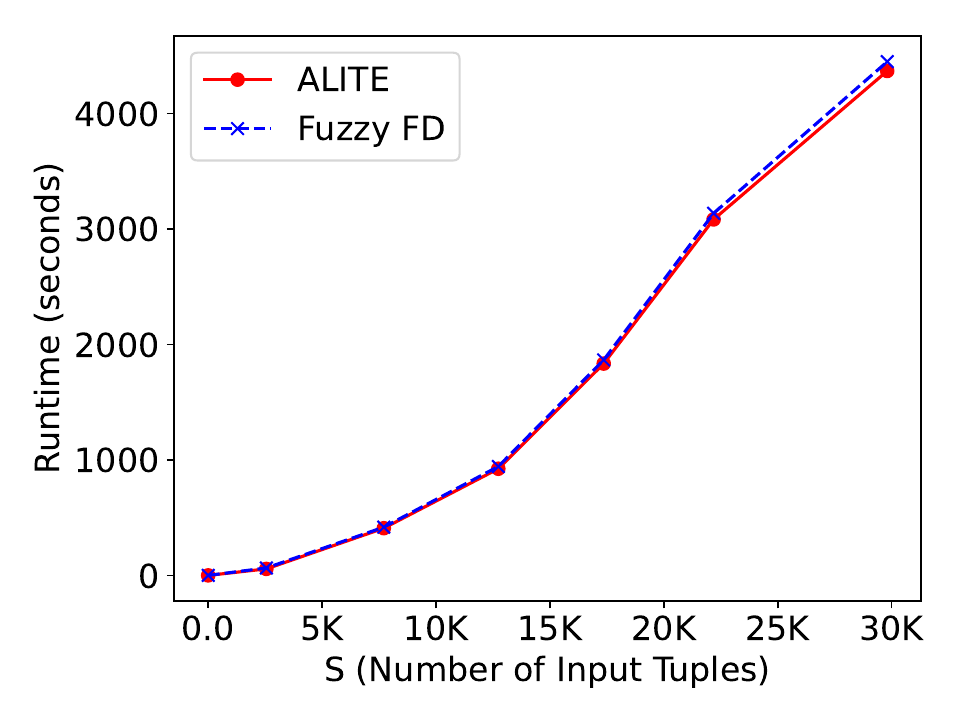}
  \caption{Runtime comparison of Regular Full Disjunction (ALITE) with Fuzzy FD in IMDB Benchmark.}
\label{fig:runtime_comparison_imdb}
\end{figure}
Now we discuss the results of our experiments.

\introparagraph{Fuzzy Matching Effectiveness}
First, we report the performance of different embedding methods in determining the fuzzy matches.
~\cref{tab:fuzzy_matching_effectiveness} shows the average performance of each embedding baseline over 31 sets of aligning columns in Auto-Join Benchmark.
It is seen that although being smaller in size (7B parameters) than the second best model Llama3 (8B parameters), Mistral outperforms all the models in terms of Precision, Recall, and $F_1$-Score. Other models, FastText, BERT, and RoBERTa, show lower performances by at least 8\% in terms of all metrics than Mistral. This shows that pre-trained embeddings of large language models can be used to embed the column values for fuzzy matches. As Mistral, even being smaller than Llama3, performs better, we use it in our system and for all our experiments.

\introparagraph{Downstreaming Task Effectiveness}
Next, we report the effectiveness of integration using Fuzzy FD over regular FD in ALITE Benchmark. Specifically, we perform entity matching over the integrated tables by fuzzy FD and regular FD (ALITE) and report the effectiveness. 
The results show that entity resolution over Fuzzy FD integration (P = 86 \%, R = 85 \%, and $F_1$ = 85 \%) 
is better than that over the regular FD (P = 79 \%, R = 83 \% and $F_1$ = 81 \%). 
Precision improves with Fuzzy FD because it eliminates the false positives that regular FD produces due to unmatched values. Additionally, Fuzzy FD's better integration of tuples provides more information for the entity matching algorithm, which as a result, retains more true positive tuples, increasing recall.

\introparagraph{Efficiency}
We compare the runtime of regular FD Operator based on ALITE~\cite{2022_khatiwada_alite} against our Fuzzy FD in IMDB Benchmark. 
For a comprehensive evaluation, we report the total number of input tuples considered for integration in the X-axis and the runtime in the Y-axis. As shown in~\cref{fig:runtime_comparison_imdb}, the lines for both methods almost overlap throughout the graph, showing that
our fuzzy FD algorithm, although needs an additional matching step, does not add additional time overhead to the regular Full Disjunction, and also increases integration effectiveness.

%% file: sections/conclusion.tex
\section{Conclusion}

We presented an extension of the Full Disjunction algorithm to integrate a set of data lake tables considering fuzzy matches between the values. Experimentally, we showed that our Fuzzy Full Disjunction algorithm is much more effective, also being as fast as the regular Full Disjunction algorithm. In the future, we will develop finetuned models to better represent the column values for further enhancing matching effectiveness.

%% file: main.bbl

\begin{thebibliography}{34}


\ifx \showCODEN    \undefined \def \showCODEN     #1{\unskip}     \fi
\ifx \showDOI      \undefined \def \showDOI       #1{#1}\fi
\ifx \showISBNx    \undefined \def \showISBNx     #1{\unskip}     \fi
\ifx \showISBNxiii \undefined \def \showISBNxiii  #1{\unskip}     \fi
\ifx \showISSN     \undefined \def \showISSN      #1{\unskip}     \fi
\ifx \showLCCN     \undefined \def \showLCCN      #1{\unskip}     \fi
\ifx \shownote     \undefined \def \shownote      #1{#1}          \fi
\ifx \showarticletitle \undefined \def \showarticletitle #1{#1}   \fi
\ifx \showURL      \undefined \def \showURL       {\relax}        \fi
\providecommand\bibfield[2]{#2}
\providecommand\bibinfo[2]{#2}
\providecommand\natexlab[1]{#1}
\providecommand\showeprint[2][]{arXiv:#2}

\bibitem[Bleiholder and Naumann(2009)]%
        {2009_bleiholder_data_fusion}
\bibfield{author}{\bibinfo{person}{Jens Bleiholder} {and} \bibinfo{person}{Felix Naumann}.} \bibinfo{year}{2009}\natexlab{}.
\newblock \showarticletitle{Data Fusion}.
\newblock \bibinfo{journal}{\emph{ACM Comput. Surv.}} \bibinfo{volume}{41}, \bibinfo{number}{1}, Article \bibinfo{articleno}{1} (\bibinfo{date}{Jan.} \bibinfo{year}{2009}), \bibinfo{numpages}{41}~pages.
\newblock
\showISSN{0360-0300}
\urldef\tempurl%
\url{https://doi.org/10.1145/1456650.1456651}
\showDOI{\tempurl}


\bibitem[Brickley et~al\mbox{.}(2019)]%
        {2019_Brickley_google_dataset_search}
\bibfield{author}{\bibinfo{person}{Dan Brickley}, \bibinfo{person}{Matthew Burgess}, {and} \bibinfo{person}{Natasha Noy}.} \bibinfo{year}{2019}\natexlab{}.
\newblock \showarticletitle{Google Dataset Search: Building a Search Engine for Datasets in an Open Web Ecosystem}. In \bibinfo{booktitle}{\emph{The World Wide Web Conference}}. \bibinfo{publisher}{ACM}, \bibinfo{pages}{1365–1375}.
\newblock
\showISBNx{9781450366748}
\urldef\tempurl%
\url{https://doi.org/10.1145/3308558.3313685}
\showDOI{\tempurl}


\bibitem[Christophides et~al\mbox{.}(2021)]%
        {2021_christophides_er_survery}
\bibfield{author}{\bibinfo{person}{Vassilis Christophides}, \bibinfo{person}{Vasilis Efthymiou}, \bibinfo{person}{Themis Palpanas}, \bibinfo{person}{George Papadakis}, {and} \bibinfo{person}{Kostas Stefanidis}.} \bibinfo{year}{2021}\natexlab{}.
\newblock \showarticletitle{An Overview of End-to-End Entity Resolution for Big Data}.
\newblock \bibinfo{journal}{\emph{{ACM} Comput. Surv.}} \bibinfo{volume}{53}, \bibinfo{number}{6} (\bibinfo{year}{2021}), \bibinfo{pages}{127:1--127:42}.
\newblock
\urldef\tempurl%
\url{https://doi.org/10.1145/3418896}
\showDOI{\tempurl}


\bibitem[Cohen et~al\mbox{.}(2006)]%
        {2006_cohen_poly_delay_fD}
\bibfield{author}{\bibinfo{person}{Sara Cohen}, \bibinfo{person}{Itzhak Fadida}, \bibinfo{person}{Yaron Kanza}, \bibinfo{person}{Benny Kimelfeld}, {and} \bibinfo{person}{Yehoshua Sagiv}.} \bibinfo{year}{2006}\natexlab{}.
\newblock \showarticletitle{Full Disjunctions: Polynomial-Delay Iterators in Action}. In \bibinfo{booktitle}{\emph{VLDB 2006}}. \bibinfo{publisher}{{ACM}}.
\newblock
\urldef\tempurl%
\url{http://dl.acm.org/citation.cfm?id=1164191}
\showURL{%
\tempurl}


\bibitem[Crouse(2016)]%
        {2016_crouse_assignment_algorithm}
\bibfield{author}{\bibinfo{person}{David~Frederic Crouse}.} \bibinfo{year}{2016}\natexlab{}.
\newblock \showarticletitle{On implementing 2D rectangular assignment algorithms}.
\newblock \bibinfo{journal}{\emph{{IEEE} Trans. Aerosp. Electron. Syst.}} \bibinfo{volume}{52}, \bibinfo{number}{4} (\bibinfo{year}{2016}), \bibinfo{pages}{1679--1696}.
\newblock
\urldef\tempurl%
\url{https://doi.org/10.1109/TAES.2016.140952}
\showDOI{\tempurl}


\bibitem[Devlin et~al\mbox{.}(2019)]%
        {2019_devlin_bert}
\bibfield{author}{\bibinfo{person}{Jacob Devlin}, \bibinfo{person}{Ming-Wei Chang}, \bibinfo{person}{Kenton Lee}, {and} \bibinfo{person}{Kristina Toutanova}.} \bibinfo{year}{2019}\natexlab{}.
\newblock \showarticletitle{BERT: Pre-training of Deep Bidirectional Transformers for Language Understanding}.
\newblock \bibinfo{journal}{\emph{ArXiv}}  \bibinfo{volume}{abs/1810.04805} (\bibinfo{year}{2019}).
\newblock


\bibitem[Dong et~al\mbox{.}(2021)]%
        {2021_dong_efficient_joinable_table_discovery}
\bibfield{author}{\bibinfo{person}{Yuyang Dong}, \bibinfo{person}{Kunihiro Takeoka}, \bibinfo{person}{Chuan Xiao}, {and} \bibinfo{person}{Masafumi Oyamada}.} \bibinfo{year}{2021}\natexlab{}.
\newblock \showarticletitle{Efficient Joinable Table Discovery in Data Lakes: {A} High-Dimensional Similarity-Based Approach}. In \bibinfo{booktitle}{\emph{37th {IEEE} International Conference on Data Engineering, {ICDE} 2021}}. \bibinfo{publisher}{{IEEE}}, \bibinfo{pages}{456--467}.
\newblock
\urldef\tempurl%
\url{https://doi.org/10.1109/ICDE51399.2021.00046}
\showDOI{\tempurl}


\bibitem[Fan et~al\mbox{.}(2023)]%
        {2023_fan_starmie}
\bibfield{author}{\bibinfo{person}{Grace Fan}, \bibinfo{person}{Jin Wang}, \bibinfo{person}{Yuliang Li}, \bibinfo{person}{Dan Zhang}, {and} \bibinfo{person}{Ren{\'{e}}e~J. Miller}.} \bibinfo{year}{2023}\natexlab{}.
\newblock \showarticletitle{Semantics-aware Dataset Discovery from Data Lakes with Contextualized Column-based Representation Learning}.
\newblock \bibinfo{journal}{\emph{{PVDLB}}} \bibinfo{volume}{16}, \bibinfo{number}{7} (\bibinfo{year}{2023}), \bibinfo{pages}{1726--1739}.
\newblock


\bibitem[Fernandez et~al\mbox{.}(2018)]%
        {2018_fernandez_aurum}
\bibfield{author}{\bibinfo{person}{Raul~Castro Fernandez}, \bibinfo{person}{Ziawasch Abedjan}, \bibinfo{person}{Famien Koko}, \bibinfo{person}{Gina Yuan}, \bibinfo{person}{Samuel Madden}, {and} \bibinfo{person}{Michael Stonebraker}.} \bibinfo{year}{2018}\natexlab{}.
\newblock \showarticletitle{Aurum: A data discovery system}. In \bibinfo{booktitle}{\emph{2018 IEEE 34th International Conference on Data Engineering (ICDE)}}. IEEE, \bibinfo{pages}{1001--1012}.
\newblock


\bibitem[Galhotra and Khurana(2020)]%
        {2020_galhotra_s3d}
\bibfield{author}{\bibinfo{person}{Sainyam Galhotra} {and} \bibinfo{person}{Udayan Khurana}.} \bibinfo{year}{2020}\natexlab{}.
\newblock \showarticletitle{Semantic Search over Structured Data}. In \bibinfo{booktitle}{\emph{CIKM 2020}}. \bibinfo{publisher}{Association for Computing Machinery}, \bibinfo{pages}{3381–3384}.
\newblock
\showISBNx{9781450368599}
\urldef\tempurl%
\url{https://doi.org/10.1145/3340531.3417426}
\showDOI{\tempurl}


\bibitem[Galindo-Legaria(1994)]%
        {1994_legaria_outerjoin_as_disjunctions}
\bibfield{author}{\bibinfo{person}{C\'{e}sar~A. Galindo-Legaria}.} \bibinfo{year}{1994}\natexlab{}.
\newblock \showarticletitle{Outerjoins as Disjunctions}. In \bibinfo{booktitle}{\emph{SIGMOD Conference 1994}}. \bibinfo{publisher}{ACM}, \bibinfo{pages}{348–358}.
\newblock
\showISBNx{0897916395}
\urldef\tempurl%
\url{https://doi.org/10.1145/191839.191908}
\showDOI{\tempurl}


\bibitem[Hai et~al\mbox{.}(2023)]%
        {2023_hai_datalake_survey}
\bibfield{author}{\bibinfo{person}{Rihan Hai}, \bibinfo{person}{Christos Koutras}, \bibinfo{person}{Christoph Quix}, {and} \bibinfo{person}{Matthias Jarke}.} \bibinfo{year}{2023}\natexlab{}.
\newblock \showarticletitle{Data Lakes: {A} Survey of Functions and Systems}.
\newblock \bibinfo{journal}{\emph{{IEEE} Trans. Knowl. Data Eng.}} \bibinfo{volume}{35}, \bibinfo{number}{12} (\bibinfo{year}{2023}), \bibinfo{pages}{12571--12590}.
\newblock
\urldef\tempurl%
\url{https://doi.org/10.1109/TKDE.2023.3270101}
\showDOI{\tempurl}


\bibitem[Hu et~al\mbox{.}(2023)]%
        {2023_hu_autotus}
\bibfield{author}{\bibinfo{person}{Xuming Hu}, \bibinfo{person}{Shen Wang}, \bibinfo{person}{Xiao Qin}, \bibinfo{person}{Chuan Lei}, \bibinfo{person}{Zhengyuan Shen}, \bibinfo{person}{Christos Faloutsos}, \bibinfo{person}{Asterios Katsifodimos}, \bibinfo{person}{George Karypis}, \bibinfo{person}{Lijie Wen}, {and} \bibinfo{person}{Philip~S. Yu}.} \bibinfo{year}{2023}\natexlab{}.
\newblock \showarticletitle{Automatic Table Union Search with Tabular Representation Learning}. In \bibinfo{booktitle}{\emph{Findings of the Association for Computational Linguistics: {ACL} 2023, Toronto, Canada, July 9-14, 2023}}. \bibinfo{publisher}{Association for Computational Linguistics}, \bibinfo{pages}{3786--3800}.
\newblock
\urldef\tempurl%
\url{https://aclanthology.org/2023.findings-acl.233}
\showURL{%
\tempurl}


\bibitem[Joulin et~al\mbox{.}(2016)]%
        {2016_joulin_fasttext}
\bibfield{author}{\bibinfo{person}{Armand Joulin}, \bibinfo{person}{Edouard Grave}, \bibinfo{person}{Piotr Bojanowski}, {and} \bibinfo{person}{Tomas Mikolov}.} \bibinfo{year}{2016}\natexlab{}.
\newblock \showarticletitle{Bag of tricks for efficient text classification}.
\newblock \bibinfo{journal}{\emph{arXiv preprint arXiv:1607.01759}} (\bibinfo{year}{2016}).
\newblock


\bibitem[Kanza and Sagiv(2003)]%
        {2003_kanza_computing_FD}
\bibfield{author}{\bibinfo{person}{Yaron Kanza} {and} \bibinfo{person}{Yehoshua Sagiv}.} \bibinfo{year}{2003}\natexlab{}.
\newblock \showarticletitle{Computing Full Disjunctions}. In \bibinfo{booktitle}{\emph{Proceedings of the Twenty-Second ACM SIGMOD-SIGACT-SIGART Symposium on Principles of Database Systems}} \emph{(\bibinfo{series}{PODS '03})}. \bibinfo{publisher}{ACM}, \bibinfo{pages}{78–89}.
\newblock
\showISBNx{1581136706}
\urldef\tempurl%
\url{https://doi.org/10.1145/773153.773162}
\showDOI{\tempurl}


\bibitem[Khatiwada et~al\mbox{.}(2023)]%
        {2023_khatiwada_santos}
\bibfield{author}{\bibinfo{person}{Aamod Khatiwada}, \bibinfo{person}{Grace Fan}, \bibinfo{person}{Roee Shraga}, \bibinfo{person}{Zixuan Chen}, \bibinfo{person}{Wolfgang Gatterbauer}, \bibinfo{person}{Ren{\'e}e~J Miller}, {and} \bibinfo{person}{Mirek Riedewald}.} \bibinfo{year}{2023}\natexlab{}.
\newblock \showarticletitle{SANTOS: Relationship-based Semantic Table Union Search}.
\newblock \bibinfo{journal}{\emph{Proc. ACM Manag. Data}} \bibinfo{volume}{1}, \bibinfo{number}{1} (\bibinfo{year}{2023}), \bibinfo{pages}{Article 9}.
\newblock
\urldef\tempurl%
\url{https://doi.org/10.1145/3588689}
\showDOI{\tempurl}


\bibitem[Khatiwada et~al\mbox{.}(2024)]%
        {khatiwada2024tabsketchfm}
\bibfield{author}{\bibinfo{person}{Aamod Khatiwada}, \bibinfo{person}{Harsha Kokel}, \bibinfo{person}{Ibrahim Abdelaziz}, \bibinfo{person}{Subhajit Chaudhury}, \bibinfo{person}{Julian Dolby}, \bibinfo{person}{Oktie Hassanzadeh}, \bibinfo{person}{Zhenhan Huang}, \bibinfo{person}{Tejaswini Pedapati}, \bibinfo{person}{Horst Samulowitz}, {and} \bibinfo{person}{Kavitha Srinivas}.} \bibinfo{year}{2024}\natexlab{}.
\newblock \showarticletitle{TabSketchFM: Sketch-based Tabular Representation Learning for Data Discovery over Data Lakes}. In \bibinfo{booktitle}{\emph{NeurIPS 2024 Third Table Representation Learning Workshop}}.
\newblock


\bibitem[Khatiwada et~al\mbox{.}(2022)]%
        {2022_khatiwada_alite}
\bibfield{author}{\bibinfo{person}{Aamod Khatiwada}, \bibinfo{person}{Roee Shraga}, \bibinfo{person}{Wolfgang Gatterbauer}, {and} \bibinfo{person}{Ren{\'{e}}e~J. Miller}.} \bibinfo{year}{2022}\natexlab{}.
\newblock \showarticletitle{Integrating Data Lake Tables}.
\newblock \bibinfo{journal}{\emph{Proc. {VLDB} Endow.}} \bibinfo{volume}{16}, \bibinfo{number}{4} (\bibinfo{year}{2022}), \bibinfo{pages}{932--945}.
\newblock
\urldef\tempurl%
\url{https://doi.org/10.14778/3574245.3574274}
\showDOI{\tempurl}


\bibitem[Koutras et~al\mbox{.}(2021)]%
        {koutras2021valentine}
\bibfield{author}{\bibinfo{person}{Christos Koutras}, \bibinfo{person}{George Siachamis}, \bibinfo{person}{Andra Ionescu}, \bibinfo{person}{Kyriakos Psarakis}, \bibinfo{person}{Jerry Brons}, \bibinfo{person}{Marios Fragkoulis}, \bibinfo{person}{Christoph Lofi}, \bibinfo{person}{Angela Bonifati}, {and} \bibinfo{person}{Asterios Katsifodimos}.} \bibinfo{year}{2021}\natexlab{}.
\newblock \showarticletitle{Valentine: Evaluating Matching Techniques for Dataset Discovery}. In \bibinfo{booktitle}{\emph{37th {IEEE} International Conference on Data Engineering, {ICDE} 2021}}. \bibinfo{publisher}{{IEEE}}, \bibinfo{pages}{468--479}.
\newblock
\urldef\tempurl%
\url{https://doi.org/10.1109/ICDE51399.2021.00047}
\showDOI{\tempurl}


\bibitem[Li et~al\mbox{.}(2021)]%
        {2021_li_auto_fuzzy_join}
\bibfield{author}{\bibinfo{person}{Peng Li}, \bibinfo{person}{Xiang Cheng}, \bibinfo{person}{Xu Chu}, \bibinfo{person}{Yeye He}, {and} \bibinfo{person}{Surajit Chaudhuri}.} \bibinfo{year}{2021}\natexlab{}.
\newblock \showarticletitle{Auto-FuzzyJoin: Auto-Program Fuzzy Similarity Joins Without Labeled Examples}. In \bibinfo{booktitle}{\emph{{SIGMOD} '21: International Conference on Management of Data, Virtual Event, China, June 20-25, 2021}}, \bibfield{editor}{\bibinfo{person}{Guoliang Li}, \bibinfo{person}{Zhanhuai Li}, \bibinfo{person}{Stratos Idreos}, {and} \bibinfo{person}{Divesh Srivastava}} (Eds.). \bibinfo{publisher}{{ACM}}, \bibinfo{pages}{1064--1076}.
\newblock
\urldef\tempurl%
\url{https://doi.org/10.1145/3448016.3452824}
\showDOI{\tempurl}


\bibitem[Li et~al\mbox{.}(2020)]%
        {2020_li_ditto}
\bibfield{author}{\bibinfo{person}{Yuliang Li}, \bibinfo{person}{Jinfeng Li}, \bibinfo{person}{Yoshihiko Suhara}, \bibinfo{person}{AnHai Doan}, {and} \bibinfo{person}{Wang{-}Chiew Tan}.} \bibinfo{year}{2020}\natexlab{}.
\newblock \showarticletitle{Deep Entity Matching with Pre-Trained Language Models}.
\newblock \bibinfo{journal}{\emph{Proc. {VLDB} Endow.}} \bibinfo{volume}{14}, \bibinfo{number}{1} (\bibinfo{year}{2020}), \bibinfo{pages}{50--60}.
\newblock
\urldef\tempurl%
\url{https://doi.org/10.14778/3421424.3421431}
\showDOI{\tempurl}


\bibitem[Liu et~al\mbox{.}(2019)]%
        {2019_liu_roberta}
\bibfield{author}{\bibinfo{person}{Yinhan Liu}, \bibinfo{person}{Myle Ott}, \bibinfo{person}{Naman Goyal}, \bibinfo{person}{Jingfei Du}, \bibinfo{person}{Mandar Joshi}, \bibinfo{person}{Danqi Chen}, \bibinfo{person}{Omer Levy}, \bibinfo{person}{Mike Lewis}, \bibinfo{person}{Luke Zettlemoyer}, {and} \bibinfo{person}{Veselin Stoyanov}.} \bibinfo{year}{2019}\natexlab{}.
\newblock \showarticletitle{RoBERTa: {A} Robustly Optimized {BERT} Pretraining Approach}.
\newblock \bibinfo{journal}{\emph{CoRR}}  \bibinfo{volume}{abs/1907.11692} (\bibinfo{year}{2019}).
\newblock
\showeprint[arXiv]{1907.11692}
\urldef\tempurl%
\url{http://arxiv.org/abs/1907.11692}
\showURL{%
\tempurl}


\bibitem[Miller(2018)]%
        {2018_miller_open_data_integration}
\bibfield{author}{\bibinfo{person}{Ren{\'{e}}e~J. Miller}.} \bibinfo{year}{2018}\natexlab{}.
\newblock \showarticletitle{Open Data Integration}.
\newblock \bibinfo{journal}{\emph{Proc. {VLDB} Endow.}} \bibinfo{volume}{11}, \bibinfo{number}{12} (\bibinfo{year}{2018}), \bibinfo{pages}{2130--2139}.
\newblock
\urldef\tempurl%
\url{https://doi.org/10.14778/3229863.3240491}
\showDOI{\tempurl}


\bibitem[Nargesian et~al\mbox{.}(2019)]%
        {2019_nargesian_data_lake_management}
\bibfield{author}{\bibinfo{person}{Fatemeh Nargesian}, \bibinfo{person}{Erkang Zhu}, \bibinfo{person}{Ren\'{e}e~J. Miller}, \bibinfo{person}{Ken~Q. Pu}, {and} \bibinfo{person}{Patricia~C. Arocena}.} \bibinfo{year}{2019}\natexlab{}.
\newblock \showarticletitle{Data Lake Management: Challenges and Opportunities}.
\newblock \bibinfo{journal}{\emph{Proc. VLDB Endow.}} \bibinfo{volume}{12}, \bibinfo{number}{12} (\bibinfo{year}{2019}), \bibinfo{pages}{1986–1989}.
\newblock
\showISSN{2150-8097}
\urldef\tempurl%
\url{https://doi.org/10.14778/3352063.3352116}
\showDOI{\tempurl}


\bibitem[Nargesian et~al\mbox{.}(2018)]%
        {2018_nargesian_tus}
\bibfield{author}{\bibinfo{person}{Fatemeh Nargesian}, \bibinfo{person}{Erkang Zhu}, \bibinfo{person}{Ken~Q. Pu}, {and} \bibinfo{person}{Ren\'{e}e~J. Miller}.} \bibinfo{year}{2018}\natexlab{}.
\newblock \showarticletitle{Table Union Search on Open Data}.
\newblock \bibinfo{journal}{\emph{Proc. VLDB Endow.}} \bibinfo{volume}{11}, \bibinfo{number}{7} (\bibinfo{year}{2018}), \bibinfo{pages}{813–825}.
\newblock
\showISSN{2150-8097}
\urldef\tempurl%
\url{https://doi.org/10.14778/3192965.3192973}
\showDOI{\tempurl}


\bibitem[Ouellette et~al\mbox{.}(2021)]%
        {2021_ouellette_ronin}
\bibfield{author}{\bibinfo{person}{Paul Ouellette}, \bibinfo{person}{Aidan Sciortino}, \bibinfo{person}{Fatemeh Nargesian}, \bibinfo{person}{Bahar~Ghadiri Bashardoost}, \bibinfo{person}{Erkang Zhu}, \bibinfo{person}{Ken~Q. Pu}, {and} \bibinfo{person}{Ren\'{e}e~J. Miller}.} \bibinfo{year}{2021}\natexlab{}.
\newblock \showarticletitle{RONIN: Data Lake Exploration}.
\newblock \bibinfo{journal}{\emph{Proc. VLDB Endow.}} \bibinfo{volume}{14}, \bibinfo{number}{12} (\bibinfo{year}{2021}), \bibinfo{pages}{2863–2866}.
\newblock
\showISSN{2150-8097}
\urldef\tempurl%
\url{https://doi.org/10.14778/3476311.3476364}
\showDOI{\tempurl}


\bibitem[Paganelli et~al\mbox{.}(2019)]%
        {2019_paganelli_parallelizing_fd}
\bibfield{author}{\bibinfo{person}{Matteo Paganelli}, \bibinfo{person}{Domenico Beneventano}, \bibinfo{person}{Francesco Guerra}, {and} \bibinfo{person}{Paolo Sottovia}.} \bibinfo{year}{2019}\natexlab{}.
\newblock \showarticletitle{Parallelizing Computations of Full Disjunctions}.
\newblock \bibinfo{journal}{\emph{Big Data Research}}  \bibinfo{volume}{17} (\bibinfo{year}{2019}), \bibinfo{pages}{18--31}.
\newblock
\showISSN{2214-5796}
\urldef\tempurl%
\url{https://doi.org/10.1016/j.bdr.2019.07.002}
\showDOI{\tempurl}


\bibitem[Papadakis et~al\mbox{.}(2020)]%
        {2021_papadakis_er_tutorial}
\bibfield{author}{\bibinfo{person}{George Papadakis}, \bibinfo{person}{Ekaterini Ioannou}, {and} \bibinfo{person}{Themis Palpanas}.} \bibinfo{year}{2020}\natexlab{}.
\newblock \showarticletitle{Entity Resolution: Past, Present and Yet-to-Come}. In \bibinfo{booktitle}{\emph{Proceedings of the 23rd International Conference on Extending Database Technology, {EDBT} 2020}}. \bibinfo{publisher}{OpenProceedings.org}, \bibinfo{pages}{647--650}.
\newblock
\urldef\tempurl%
\url{https://doi.org/10.5441/002/edbt.2020.85}
\showDOI{\tempurl}


\bibitem[Rajaraman and Ullman(1996)]%
        {1996_ullman_outer_join_gamma_cycles}
\bibfield{author}{\bibinfo{person}{Anand Rajaraman} {and} \bibinfo{person}{Jeffrey~D. Ullman}.} \bibinfo{year}{1996}\natexlab{}.
\newblock \showarticletitle{Integrating Information by Outerjoins and Full Disjunctions (Extended Abstract)}. In \bibinfo{booktitle}{\emph{PODS 1996}}. \bibinfo{publisher}{ACM}.
\newblock
\showISBNx{0897917812}


\bibitem[Ravat and Zhao(2019)]%
        {2019_ravat_data_lakes_trends}
\bibfield{author}{\bibinfo{person}{Franck Ravat} {and} \bibinfo{person}{Yan Zhao}.} \bibinfo{year}{2019}\natexlab{}.
\newblock \showarticletitle{Data Lakes: Trends and Perspectives}. In \bibinfo{booktitle}{\emph{Database and Expert Systems Applications - 30th International Conference, {DEXA} 2019, Linz, Austria, August 26-29, 2019, Proceedings, Part {I}}} \emph{(\bibinfo{series}{Lecture Notes in Computer Science}, Vol.~\bibinfo{volume}{11706})}, \bibfield{editor}{\bibinfo{person}{Sven Hartmann}, \bibinfo{person}{Josef K{\"{u}}ng}, \bibinfo{person}{Sharma Chakravarthy}, \bibinfo{person}{Gabriele Anderst{-}Kotsis}, \bibinfo{person}{A~Min Tjoa}, {and} \bibinfo{person}{Ismail Khalil}} (Eds.). \bibinfo{publisher}{Springer}, \bibinfo{pages}{304--313}.
\newblock
\urldef\tempurl%
\url{https://doi.org/10.1007/978-3-030-27615-7\_23}
\showDOI{\tempurl}


\bibitem[Su et~al\mbox{.}(2006)]%
        {su2006holistic}
\bibfield{author}{\bibinfo{person}{Weifeng Su}, \bibinfo{person}{Jiying Wang}, {and} \bibinfo{person}{Frederick~H. Lochovsky}.} \bibinfo{year}{2006}\natexlab{}.
\newblock \showarticletitle{Holistic Schema Matching for Web Query Interfaces}. In \bibinfo{booktitle}{\emph{Advances in Database Technology - {EDBT} 2006, 10th International Conference on Extending Database Technology, Proceedings}} \emph{(\bibinfo{series}{Lecture Notes in Computer Science}, Vol.~\bibinfo{volume}{3896})}. \bibinfo{publisher}{Springer}, \bibinfo{pages}{77--94}.
\newblock
\urldef\tempurl%
\url{https://doi.org/10.1007/11687238\_8}
\showDOI{\tempurl}


\bibitem[Zhu et~al\mbox{.}(2019)]%
        {2019_zhu_josie}
\bibfield{author}{\bibinfo{person}{Erkang Zhu}, \bibinfo{person}{Dong Deng}, \bibinfo{person}{Fatemeh Nargesian}, {and} \bibinfo{person}{Ren{\'{e}}e~J. Miller}.} \bibinfo{year}{2019}\natexlab{}.
\newblock \showarticletitle{{JOSIE:} Overlap Set Similarity Search for Finding Joinable Tables in Data Lakes}. In \bibinfo{booktitle}{\emph{SIGMOD Conference 2019}}. \bibinfo{publisher}{{ACM}}, \bibinfo{pages}{847--864}.
\newblock
\urldef\tempurl%
\url{https://doi.org/10.1145/3299869.3300065}
\showDOI{\tempurl}


\bibitem[Zhu et~al\mbox{.}(2017)]%
        {2017_zhu_autojoin}
\bibfield{author}{\bibinfo{person}{Erkang Zhu}, \bibinfo{person}{Yeye He}, {and} \bibinfo{person}{Surajit Chaudhuri}.} \bibinfo{year}{2017}\natexlab{}.
\newblock \showarticletitle{Auto-Join: Joining Tables by Leveraging Transformations}.
\newblock \bibinfo{journal}{\emph{Proc. {VLDB} Endow.}} \bibinfo{volume}{10}, \bibinfo{number}{10} (\bibinfo{year}{2017}), \bibinfo{pages}{1034--1045}.
\newblock
\urldef\tempurl%
\url{https://doi.org/10.14778/3115404.3115409}
\showDOI{\tempurl}


\bibitem[Zhu et~al\mbox{.}(2016)]%
        {DBLP:journals/pvldb/ZhuNPM16}
\bibfield{author}{\bibinfo{person}{Erkang Zhu}, \bibinfo{person}{Fatemeh Nargesian}, \bibinfo{person}{Ken~Q. Pu}, {and} \bibinfo{person}{Ren{\'{e}}e~J. Miller}.} \bibinfo{year}{2016}\natexlab{}.
\newblock \showarticletitle{{LSH} Ensemble: Internet-Scale Domain Search}.
\newblock \bibinfo{journal}{\emph{Proc. {VLDB} Endow.}} \bibinfo{volume}{9}, \bibinfo{number}{12} (\bibinfo{year}{2016}), \bibinfo{pages}{1185--1196}.
\newblock
\urldef\tempurl%
\url{https://doi.org/10.14778/2994509.2994534}
\showDOI{\tempurl}


\end{thebibliography}
